\documentclass[prl,twocolumn,nofootinbib,superscriptaddress]{revtex4}
\usepackage{calrsfs}
\usepackage{ulem}
\usepackage{graphics}
\usepackage{epstopdf}
\usepackage{bm}
\usepackage{bbm}
\usepackage{amssymb}
\usepackage{epstopdf}
\usepackage{graphicx}
\usepackage[caption=false]{subfig}
\usepackage{amsmath}
\usepackage{color}
\usepackage{array}

\begin{document}

\title{Ray-tracing with quantum correlated photons to image a 3D scene}

\author{Yingwen \surname{Zhang}}
\affiliation{National Research Council of Canada, 100 Sussex Drive, Ottawa ON Canada, K1A0R6}

\author{Antony \surname{Orth}}
\affiliation{National Research Council of Canada, 100 Sussex Drive, Ottawa ON Canada, K1A0R6}

\author{Duncan \surname{England}}
\email{Duncan.England@nrc-cnrc.gc.ca}
\affiliation{National Research Council of Canada, 100 Sussex Drive, Ottawa ON Canada, K1A0R6}

\author{Benjamin \surname{Sussman}}
\affiliation{National Research Council of Canada, 100 Sussex Drive, Ottawa ON Canada, K1A0R6}
\affiliation{Department of Physics, University of Ottawa, Ottawa, Ontario, K1N 6N5, Canada}

\begin{abstract}
To capture the 3D information of a scene, conventional techniques often require multiple 2D images of the scene to be captured from different perspectives. In this work we demonstrate the reconstruction of a scene's 3D information through ray-tracing using quantum correlated photon pairs. By capturing the two photons in different image planes using time-tagging cameras and taking advantage of the position, momentum and time correlation of the photons, the photons’ propagation trajectory can be reconstructed. With this information on every photon pair, we were able to demonstrate refocusing, depth of field adjustment and parallax visualization of a 3D scene. With future camera advancements, this technique could achieve a much higher momentum resolution than conventional techniques thus giving larger depth of field and more viewing angles. The high photon correlation and low photon flux from a quantum source also makes the technique well suited for 3D imaging of light sensitive samples.
\end{abstract}	

\maketitle


In conventional optical imaging, only the 2D spatial distribution of light on the camera/detector is recorded. By capturing multiple 2D images from different perspectives, both position and angular information of the light rays can be gained, thus, 3D information of the scene can be reconstructed. This allows capabilities such as refocusing, depth of field adjustments and parallax viewing of the scene to be performed, all in post-processing. One way to achieve this is through a moving light source, such as in Fourier Ptychography~\cite{Zheng2011,Zheng2013}, whereby a sample is scanned at different angles by light emitted from individual elements of a LED array. Alternatively, with a fixed light source, cameras can be placed at different locations/angles relative to the scene, such as in axially distributed sensing~\cite{Schulein2009}, where the camera is moved relative to the subject or vice versa. An alternative technique that requires no moving parts, and only a single camera, is known as plenoptic or light field imaging~\cite{Aldelson1992,raytrix}. In this approach a microlens array is placed one focal length away from a CCD sensor, each microlens illuminates a subset of the pixels in the CCD. By knowing which lens the light ray enters, and onto which pixel it subsequently focuses, one can obtain both position and angular information of the light ray respectively. Plenoptic imaging is akin to imaging a scene simultaneously with an array of cameras thus requires no scanning or moving parts. This class of 3D imaging techniques of placing camera(s) at different angles relative to the scene is also known as Integral imaging \cite{Xiao2013,Martinez2018}. A multitude of research in applications using these techniques has been performed in recent decades, to name a few, this includes target recognition~\cite{Matoba2001,Kishk2003}, microscopy~\cite{Jang2004,Levoy2006,Zheng2013,Prevedel2014}, particle tracking~\cite{Fahringer2015,Hall2016}, wavefront sensing~\cite{Lv2016,Wu2016,Wu20162}, and microendoscopy~\cite{Orth2019}.


Due to the way these conventional techniques are performed, they tend to have a relatively low angular resolution compared to their position resolution as a result of the number of viewing angles limited by either the number of light sources or camera placement locations. And in the case of Plenoptic imaging, position resolution has to be sacrificed for angular resolution. To overcome these limitations, the use of temporally and spatially correlated classical or quantum light has been proposed~\cite{DAngelo2016,Pepe2016,DiLena2018}. Where one beam illuminates a scene to obtain position information, and the momentum/angular information of the correlated partner beam is measured on a separate sensor. In this way both position and momentum can be measured with high resolution allowing larger depth of field and more parallax viewing angles. This technique, termed \textit{Correlation Plenoptic Imaging}, has been demonstrated using weakly correlated thermal light~\cite{Pepe2017,DiLena2020}, but further advantages have been predicted using highly correlated quantum light~\cite{Pepe2016}. A similar approach using quantum correlated light has also been used to demonstrate Fourier Ptychography for amplitude and phase imaging~\cite{Aidukas2019}.

Based on the proposal in~\cite{Pepe2016}, here, we demonstrate the reconstruction of a scene's 3D information through ray-tracing using quantum correlated photon pairs for which we term the technique \textit{Quantum Correlated Ray-Tracing Imaging} (QCRTI). As shall be seen, our technique shares similarities with both Plenoptic imaging and Fourier Ptychography, but also has important differences. QCRTI uses quantum correlated photon pairs generated through the process of spontaneous parametric down conversion (SPDC) with the aid of a time tagging camera, capable of time tagging every photon detection with nanosecond precision. By imaging one photon in the crystal's Fourier plane and it's partner photon in the crystal image plane, then taking advantage of the strong time, position and momentum correlation properties of the SPDC photon pairs, we were able to trace the propagation trajectory of all the detected photon pairs. Just as in Plenoptic imaging, no scanning or moving parts are required to measure the photon trajectories, though here, the position and momentum measurement comes from the quantum correlation rather than through a microlens array. QCRTI is also similar to Fourier Ptychography in the sense that each photon is illuminating the scene from a different angle, but the illumination randomly changes due to the stochastic nature of quantum light, thus separate light sources are not required. In this proof-of-principle demonstration, we show that QCRTI can achieve various 3D imaging capabilities, such as refocusing, depth-of-field adjustments and parallax viewing of a scene all at a very low photon flux of $\sim1000$ photons per second per pixel (or $5\times10^{-11}$\,W/cm$^2$ on the sample). 

\begin{figure}
    \centering
    \includegraphics[width=1\linewidth]{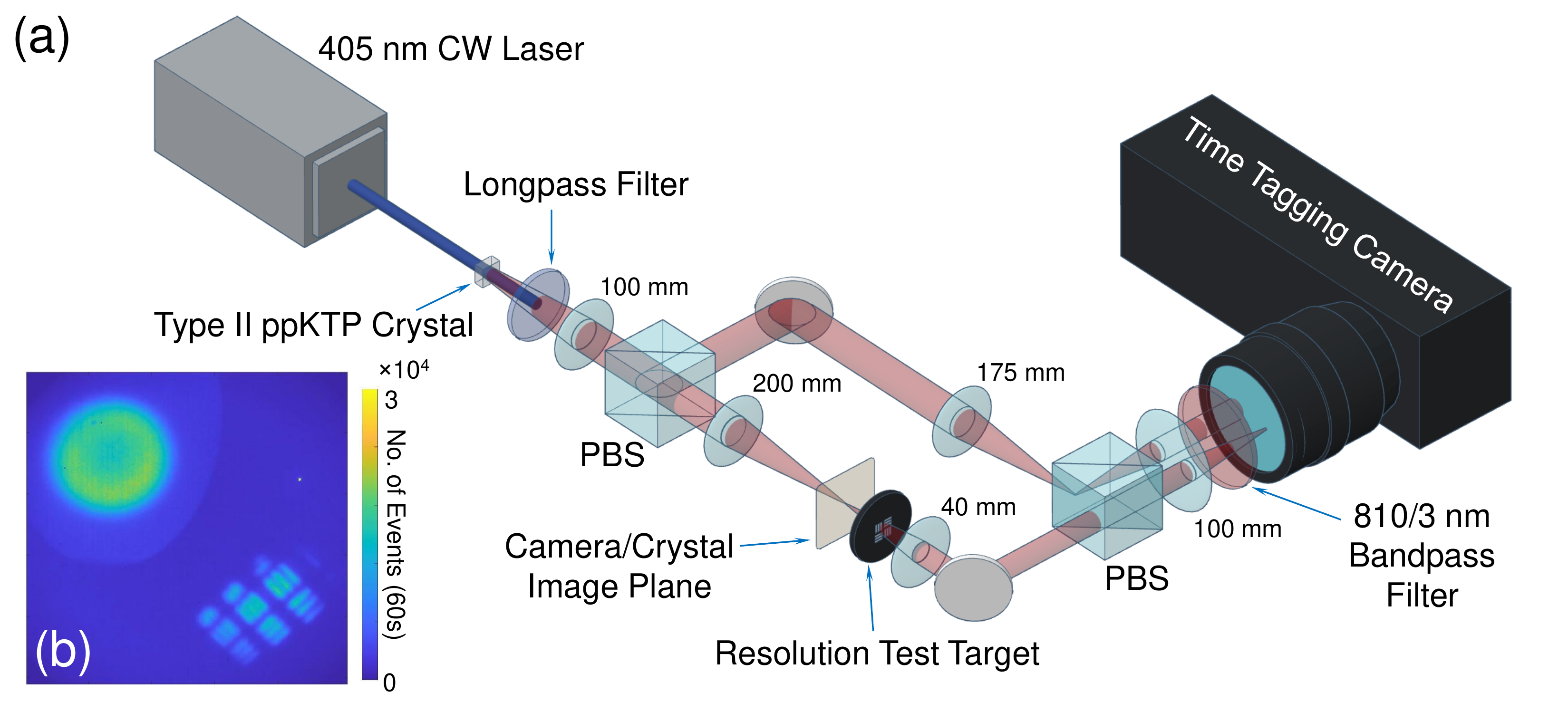}
    \caption{(a) Schematic of the QCRTI setup. (b) Raw image, before any post processing, captured on the TPX3CAM with the resolution target placed 4\,mm off focus.}
    \label{Psetup}
\end{figure}
\textit{Experimental Setup and Method} - The experimental setup is shown in Fig.~\ref{Psetup}. A 405\,nm continuous wave (CW) laser is used to pump a 1\,mm thick Type II ppKTP crystal to produce 810\,nm, orthogonally polarized photon pairs correlated in time, position and momentum through the process of spontaneous parametric down-conversion (SPDC). A longpass spectral filter placed directly after the crystal is used to block out the 405\,nm pump and let through the 810\,nm SPDC photons. As the SPDC photons are emitted from the crystal in a divergent manner, a 100\,mm lens is used to collimate the SPDC beam. Thereafter a polarizing beamsplitter (PBS) is used to split the photon pairs into separate paths, in which the sample to be imaged is placed in one of the path. The two photon beams are then recombined, but slightly displaced, by a second PBS just before the time tagging camera (TPX3CAM~\cite{Nomerotski2019,ASI}) such that each beam will be imaged onto different locations of the camera. In one path, through two magnifying 4f imaging systems, the crystal plane is first imaged onto the location at which the sample is to be placed and then onto the camera, with the beam spot magnified by 5 times. In the other path, 3 lenses are used such that the Fourier plane of the crystal is imaged onto the camera with a slightly demagnified beam spot. Ideally, two cameras can be used, one for each beam spot, to achieve higher position and momentum resolution.

A thin (1\,mm) nonlinear crystal is used to reduce the uncertainty in the depth at which the SPDC photons are generated inside the crystal, this will improve the photon pair's position correlation. Also, since the SPDC photon pairs' wavelength are not necessary degenerate, uncertainties in their momentum correlation will be introduced. To reduce this uncertainty and also to reduce background light, a 810/3\,nm spectral bandpass filter is placed just before the camera.   

The power of the UV laser has been attenuated down to 20\,mW, generating a total of approximately $1.3\times10^7$ photon pairs per second or an average photon flux of $\sim1000$\ photons per pixel per second (after accounting for the system quantum efficiency of 4\%). Increasing the pump beam power will produce more SPDC photons thereby speeding up the data acquisition process, however, at the cost of reduced spatio-temporal correlation between the photons thereby introducing more noise in the images. This is due to the increased likelihood of multi-photon pair production in the crystal during a detection time window, thus causing the incorrect photons being identified as a pair during time correlation analysis. Inversely, by reducing the pump power, one can reduce the image noise but at the cost of longer data acquisition time. 

In post processing, a virtual circular aperture is placed around each beam in the images in order to limit time correlation analysis to just the photons within the two beams thus reducing noise. Time correlation analysis is then performed on all photons detected between the two selected beam spots to identify the SPDC photon pairs. Finally, by utilizing the property that the photon pairs are perfectly correlated in position and anti-correlated in momentum, and the fact that all the optical elements and distances between them are known, we can make use of the Klyshko picture~\cite{Klyshko1988} to backtrack the propagation path of one photon from its near field to the far field of the other or vice versa, treating the crystal as a mirror, and gain both the position and momentum information of the photon pairs. Since the system is still mostly paraxial (the maximum illumination angle on our sample is $\sim0.85$\,deg), the trajectory of the photons can be determined through a simple ray transfer matrix analysis
\begin{equation}
    \begin{bmatrix}
    \overrightarrow{r_2} \\
    \overrightarrow{\theta_2} 
    \end{bmatrix}
    =
    \begin{bmatrix}
    A & B \\
    C & D
    \end{bmatrix}    
    \begin{bmatrix}
    \overrightarrow{r_1} \\
    \overrightarrow{\theta_1} 
    \end{bmatrix},
\label{ABCDmatrix}
\end{equation}
where $\overrightarrow{r_1}$ and $\overrightarrow{r_2}$ are the positions of each photon pair detected at the two camera image planes, $\overrightarrow{\theta_1}$ and $\overrightarrow{\theta_2}$ are the angles at which the photons hits each plane and the ray transfer (ABCD) matrix is determined by the optical components placed between the two image planes. Since $\overrightarrow{r_1}$ and $\overrightarrow{r_2}$ are known from the raw image data taken by the two cameras and the ABCD matrix is also known, $\overrightarrow{\theta_1}$ and $\overrightarrow{\theta_2}$ can be easily determined from Eq.~\ref{ABCDmatrix}. Thus, the full propagation trajectory of every detected photon pair is known, and with this information, one can perform in post-processing, the refocusing, depth of field adjustments and parallax viewing of the scene. More details on the camera and data processing can be found in the supplementary materials.

\begin{figure}
    \centering
    \includegraphics[width=1\linewidth]{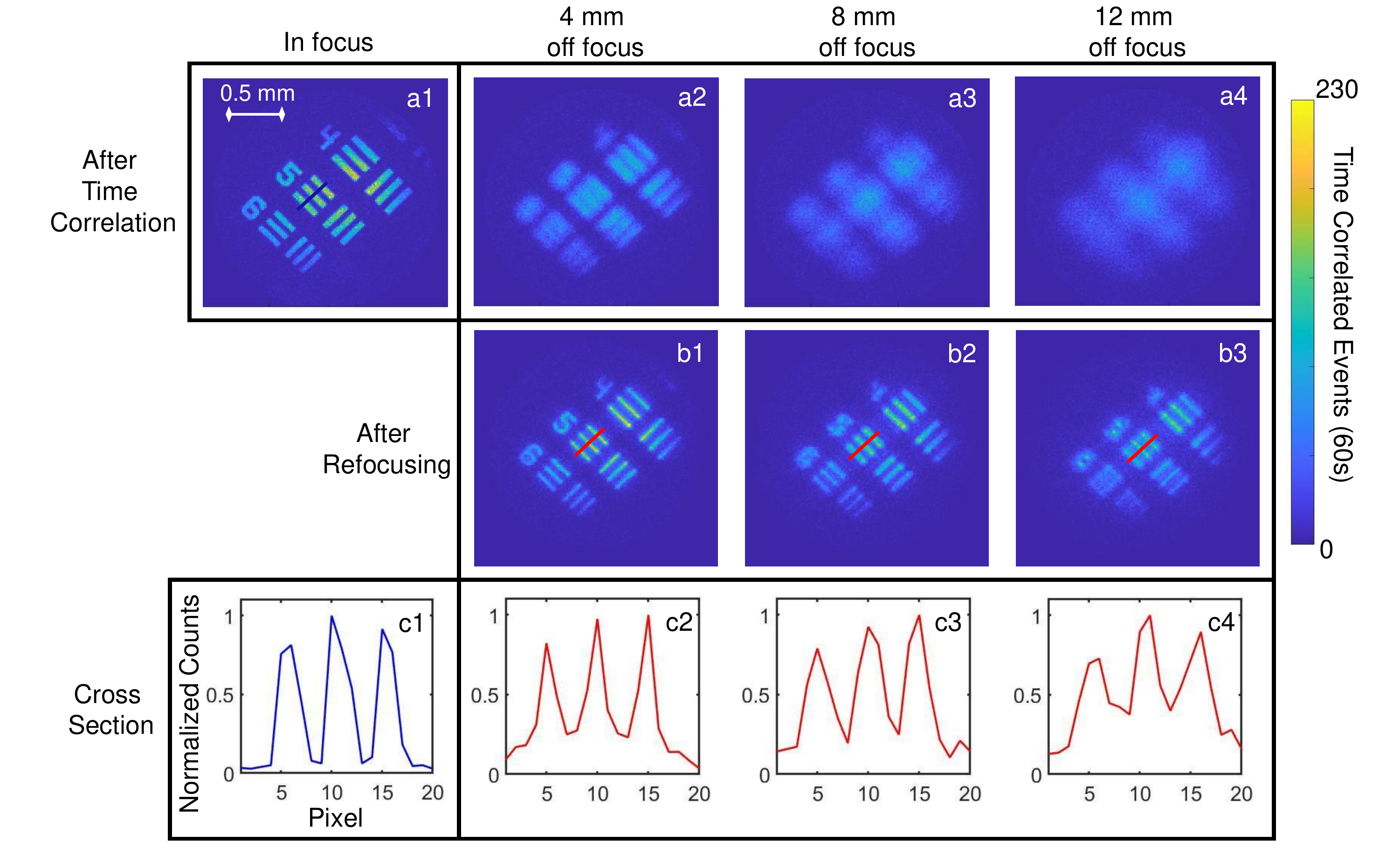}
    \caption{Demonstration of image refocusing in QCRTI using a 1951 USAF resolution test target placed at different distances from the focal plane of the imaging lenses. The line sets has a spacing of $\sim5$ line pairs per mm. The images in the top row, from a1-a4, are the images obtained after performing only time correlation measurements on the detected photons, with a1 the ground truth of the resolution test target when placed in focus. The second row, from b1-b3, are images obtained after refocusing is performed in post-processing on a2-a4 respectively. The third row, shows the cross-section through the blue line in a1 (c1) and red lines in b1-b3 (c2-c4 respectively). Each image is taken over a period of 60 seconds. The diameter of the virtual aperture seen in the raw images a1-a4 is 130 pixels.}
    \label{Prefocus}
\end{figure}

\textit{Experimental Results} -  In Fig.~\ref{Prefocus}, the refocusing capability of QCRTI is demonstrated on images taken of a 1951 USAF resolution test target placed at different distances from the focal plane of the imaging lenses. It can be seen that when the target is placed 4\,mm away from the focus, in a conventional image (Fig.~\ref{Prefocus}\,a2) the lines in the resolution target can no longer be clearly observed, however, through QCRTI they can be brought back into focus (Fig.~\ref{Prefocus}\,b1) with good image sharpness. Even at extreme distances, where features of the resolution target can no longer be identified through conventional imaging (Fig.~\ref{Prefocus}\,a4), QCRTI can still bring the image mostly back into focus (Fig.~\ref{Prefocus}\,b3). As can be seen, the sharpness of the refocusing deteriorates when the target is placed further away from the focus, this may be due to a multitude of factors such as limited position and momentum resolution, imperfect position correlation caused by the finite thickness of the crystal introducing uncertainties over the depth at which the photon pairs are generated, and lastly, imperfect momentum anti-correlation caused by the photon pairs having non-degenerate wavelength/energy and pump beam divergence at the crystal. The affect these factors have on QCRTI will require further, more detailed investigation in the future and are not within the scope of this work.

\begin{figure}
    \centering
    \includegraphics[width=1\linewidth]{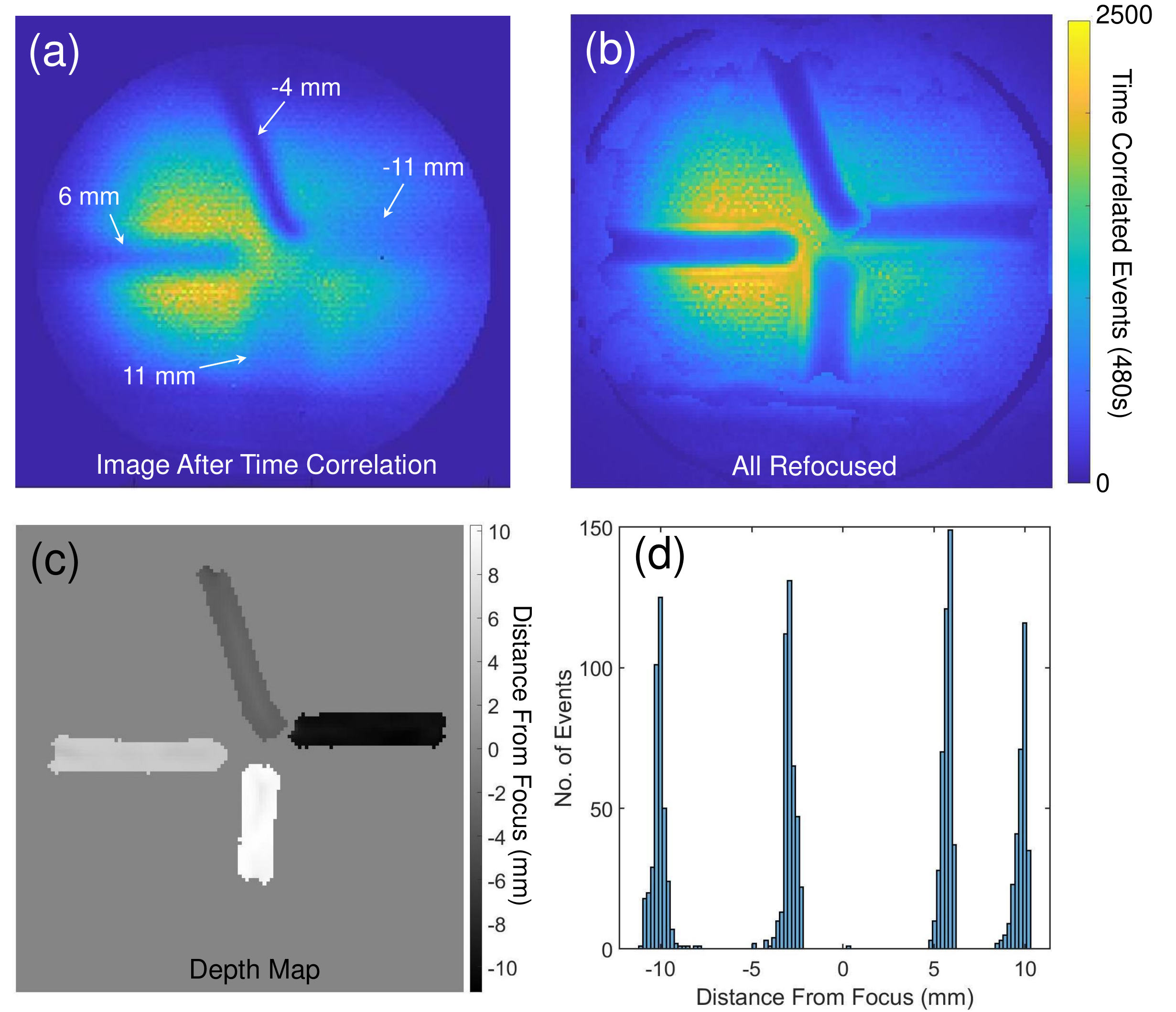}
    \caption{Images demonstrating the refocusing and depth reconstruction of 4 thin wires of 0.15\,mm diameter placed at different distances from the focus. (a) Raw image of the 4 wires after performing time correlation measurements with the manually measured distance from focus shown on image. (b) All wires brought to focus through post processing. (c) Depth map of the wires with the histogram of the depth shown in (d).}
    \label{DepthMap}
\end{figure}

In Fig.~\ref{DepthMap}, we show the refocusing and a depth map of 4 thin wires of 0.15\,mm diameter placed at different depths within the beam. Fig.~\ref{DepthMap}\,a shows the traditional image of the scene after time correlation, without momentum filtering.  Using the photon momentum information, this image is refocused into a stack of images, each corresponding to a different focus plane. The depth map in Fig.~\ref{DepthMap}\,c  is obtained by choosing the sharpest refocused plane for each pixel in the image~\cite{Nayar1994}.  Good agreement is achieved between the reconstructed depth and the manually measured depth (Fig.~\ref{DepthMap}\,d). The slight discrepancy is likely due to inaccuracies in manually measuring with a ruler, the wire locations and the distances between the optical components (lenses, mirrors etc.). This measurement inaccuracy however, does not affect the image sharpness in refocusing, as any inaccuracies here is a global effect affecting all photons in the same manner, thus, it will only introduce errors in parameters such as the size and position of the sample.  From this depth map, an all-in-focus image is constructed in Fig.~\ref{DepthMap}\,b by modifying the refocus position to the depth map value in Fig.~\ref{DepthMap}\,c on a per-pixel basis. More details on how the depth mapping is performed can be found in the supplementary material.

\begin{figure}
    \centering
    \includegraphics[width=1\linewidth]{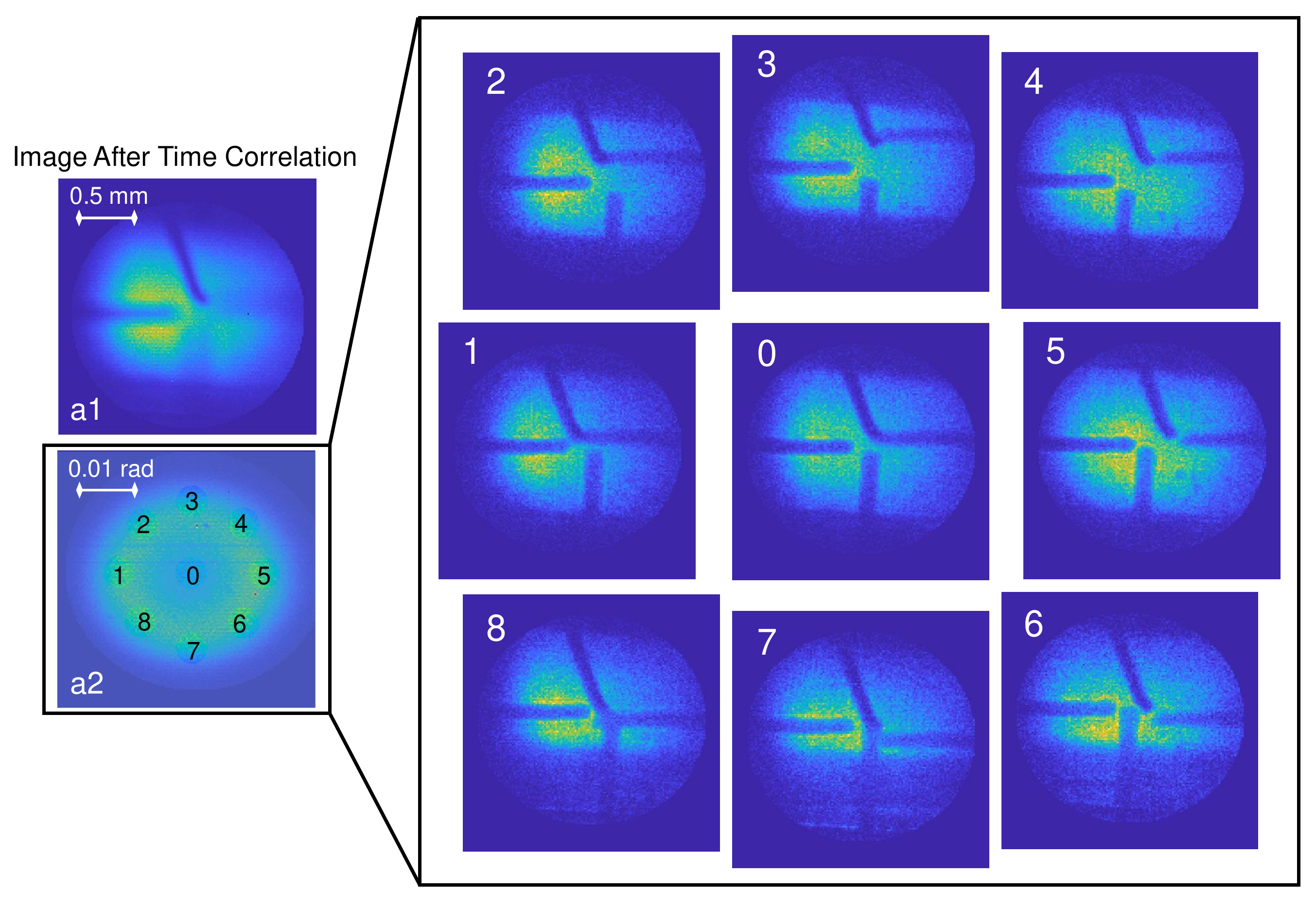}
    \caption{Images demonstrating parallax visualization and depth-of-field enhancement on 4 thin wires placed at different depth from the focus. a1 and a2 shows the raw images taken in the crystal plane and it's Fourier plane respectively after time correlation analysis. To demonstrate parallax viewing, correlation events between the full crystal plane and 9 sub-regions of the Fourier plane (as indicated in a2) were selected out. Each sub-region of a2 corresponds to a different viewing angle of the wires, these are displayed as a set of 9 images, labeled 0-8, displayed on the right, each corresponding to the 9 labeled sub-regions (illumination/viewing angles) in a2 respectively. Due to the small diameter, i.e. small momentum spread, of each of the sub-regions, an increase in the depth-of-field is also achieved. The diameter of each sub-region is 20 pixels wide, with the diameter of the virtual aperture around the full beam at 130 pixels. The raw image data was acquired over 8 minutes to ensure enough photons are accumulated within each sub-region. Smaller sub-regions can potentially be used but will require substantially longer data acquisition times.}
    \label{Parallax}
\end{figure}

The demonstration of depth-of-field enhancement and parallax visualization on the 4 wires is shown in Fig.~\ref{Parallax}. By selecting coincidence events from a smaller sub-region in the Fourier plane, we can limit the allowed photon momentum, thus adjusting, in post-processing, the imaging depth-of-field in the corresponding crystal image plane. Choosing small sub-regions off centre from the optical axis results in the Fourier plane beam, results in viewing the wires in the crystal image plane from different angles thus allowing parallax visualization to be performed. 

It is important to note that QCRTI does not necessarily require the sample to be placed near the image plane of the crystal, one can have the sample placed in the crystal's Fourier plane and still perform QCRTI. As the beam is more collimated and larger in the Fourier plane, it will be more suited for imaging larger objects with more depth. Demonstration of this is shown in the supplementary materials. Videos demonstrating the post-processing of refocusing, depth of field adjustments and parallax viewing are also shown in the supplementary materials.

\textit{Conclusion} - To conclude, in this proof of concept demonstration of QCRTI, we have demonstrated some of the major capabilities of 3D imaging, including refocusing, depth of field adjustments and parallax visualization of an object/scene. QCRTI shares similarities with both Plenoptic imaging and Fourier Ptychography. QCRTI requires no scanning or moving parts to capture information on the photon trajectories, in which it shares similarity with Plenoptic imaging. However, in the Klysko picture, where the photon trajectory is backtracked from the crystal's Fourier plane to the image plane, the similarities lies with Fourier Ptychography in which each pixel of the Fourier plane camera can be treated as a light source that randomly emits photons. 

QCRTI exhibits conceptual advantages over conventional 3D imaging, and this initial demonstration takes significant steps toward realising these advantages. In particular the momentum/angle and position of the photon can both be measured with high resolution. In conventional techniques, the position is typically measured by a high resolution camera, but angular resolution is limited either by the number of camera positions (integral imaging), the number of light sources (Fourier Ptychography), or the pitch of the microlens array (plenoptic imaging). In QCRTI, the momentum resolution is only limited by the number of pixels in camera placed in the fourier plane and the photons' degree of momentum correlation, also, unlike in plenoptic imaging, one does not need to sacrifice imaging pixels to measure angular information. Here around $10^4$ pixels were used, but with new high-resolution time-tagging cameras~\cite{Morimoto2020} this could increase to $10^6$ in the near future. Therefore, with a single quantum light source, one could effectively illuminate a 3D scene from millions of different angles, a goal that is impractical with conventional techniques.

Since QCRTI is based on quantum photon correlations, it also gains some of the advantages of quantum correlation imaging over classical correlation imaging techniques. As a result of the sub-Poissonian photon statistics of the SPDC photons giving a much higher second-order photon correlations compared to classical sources, a much lower background noise can be theoretically obtained using SPDC photons in correlation imaging under low illumination conditions~\cite{Berchera_2019,Moreau2019}. This makes QCRTI potentially well suited for imaging light sensitive samples, and we expect other advantages afforded by quantum enhancements~\cite{Genovese2016} to manifest in improved imaging in future work.


\section*{Acknowledgements}
The authors are grateful to Philip Bustard, Frédéric Bouchard, Khabat Heshami, Denis Guay, and Doug Moffatt for technical support and stimulating discussion. This work was partly supported by Defence Research and Development Canada.









\bibliographystyle{unsrt}
\bibliography{QCRTIref}

\end{document}


\title{Ray-Tracing with Quantum Correlated Photons to Image a 3D Scene - Supplementary Materials}

\author{Yingwen \surname{Zhang}}
\affiliation{National Research Council of Canada, 100 Sussex Drive, Ottawa ON Canada, K1A0R6}

\author{Antony \surname{Orth}}
\affiliation{National Research Council of Canada, 100 Sussex Drive, Ottawa ON Canada, K1A0R6}

\author{Duncan \surname{England}}
\email{Duncan.England@nrc-cnrc.gc.ca}
\affiliation{National Research Council of Canada, 100 Sussex Drive, Ottawa ON Canada, K1A0R6}

\author{Benjamin \surname{Sussman}}
\affiliation{National Research Council of Canada, 100 Sussex Drive, Ottawa ON Canada, K1A0R6}
\affiliation{Department of Physics, University of Ottawa, Ottawa, Ontario, K1N 6N5, Canada}

\maketitle

\section{Camera Operations}
Single photon sensitivity of the TPX3CAM is provided by the attached image intensifier which has a quantum efficiency of approximately 20\% at around 800\,nm~\cite{cricket} and a dark count rate of $\sim 100$\,kHz/cm$^2$. The overall quantum efficiency of the setup is $\sim4$\%, a combined effect from the efficiency of the image intensifier and all the optical elements used in the setup. As a single photon is converted into a flash of light by the image intensifier, a cluster of pixels will often be illuminated on the camera. Such a cluster has to be regrouped into a single event through a detection and centroiding algorithm~\cite{Ianzano2020}. Contrary to expectations, this step in fact introduces little to no uncertainties on the photon position. With an average cluster diameter of $~7$ pixels, the centroid of the cluster can actually be pinpointed to a smaller area than the pixel size on the camera. There has been recent demonstrations of using the centroiding to enhance camera resolution to beyond that limited by the number of pixels~\cite{Kim2020}. Due to the signal strength (photon flux) dependence of the camera's discrimination threshold crossover time, a timing correction must also be performed~\cite{Ianzano2020}, after which a timing resolution of approximately 6\,ns can be achieved. 

\section{Data Processing}
Data processing in QCRTI is much different from conventional classical methods based on processing of multiple 2D images. With our time-tagging camera, raw data is presented in a table, listing the detection time and pixel number (position) for each detected photon. A typical two photon time correlation analysis is performed between the photons detected in the two beams. Here, the difference between the detection time of the photons are compared, and a histogram of this time difference is generated. If correlation is present between the photons, a peak will be observed in the histogram. All photon pairs within the correlation peak are pulled out of the raw data (using a gating time of 20\,ns) and entered into a new table listing the pixel number for each pair of photon. 

The photon pairs' pixel number is then converted to a distance from their respective beam centre, giving the $\overrightarrow{r_1}$ and $\overrightarrow{r_2}$ in Eq.\,1 of the main text. Then using the ABCD matrix as determined by the optical components used in our setup, we can determine the angles $\overrightarrow{\theta_1}$ and $\overrightarrow{\theta_2}$ for all detected photon pairs thus giving us their propagation trajectory.

Now knowing all four parameters, $\overrightarrow{r_1}$, $\overrightarrow{r_2}$, $\overrightarrow{\theta_1}$ and $\overrightarrow{\theta_2}$, refocusing can simply be performed through adjusting the ABCD matrix to reflect moving of the final imaging plane. Depth of field and parallax viewing angle adjustments can be done by selecting out photon pairs detected within a virtual spatial filter placed in the Fourier plane beam. 

\section{Depth Mapping}
Depth mapping in Fig.\,3 on the main text is performed using a shape from focus method\cite{Nayar1994,Orth2019}.  First, the scene is computationally refocused over a range of $z = -20$\,mm to $+20$\,mm in steps of $1$\,mm to create a focal stack with 41 slices.  Next, the sum-modified-Laplacian $ML(x,y,z)= (\frac{\partial^2 I(x,y,z)}{\partial x^2})^2 + (\frac{\partial^2 I(x,y,z)}{\partial y^2})^2$  is calculated for every pixel in the focus stack.  The sum-modified-Laplacian is a heuristic sharpness metric that indicates whether local region of an image is in focus.  The depth of the scene $D(x_p,y_p)$ at each $(x,y)$ point is found by locating the z-value for which $ML(x,y,z)$ reaches its maximum value.  For further depth precision, a 1-D Gaussian is fit to $ML(x,y,z)$ peak at the initial estimated depth value for each pixel.  The all-in-focus image in Fig.\,3b is obtained by refocusing the dataset to the estimated depth $D(x,y)$ at each $(x,y)$ position independently.  For visualization purposes, the 4 wires are segmented from the background by performing adaptive thresholding over the all-in-focus image (Fig.\,3\,b).  The depth for pixel regions corresponding to the background are set to 0 in Fig.\,3c, and are removed from the histogram in Fig.\,3\,d.

\section{Quantum Correlated Ray-Tracing Imaging in the Fourier plane}
\begin{figure}[h]
    \centering
    \includegraphics[width=1\linewidth]{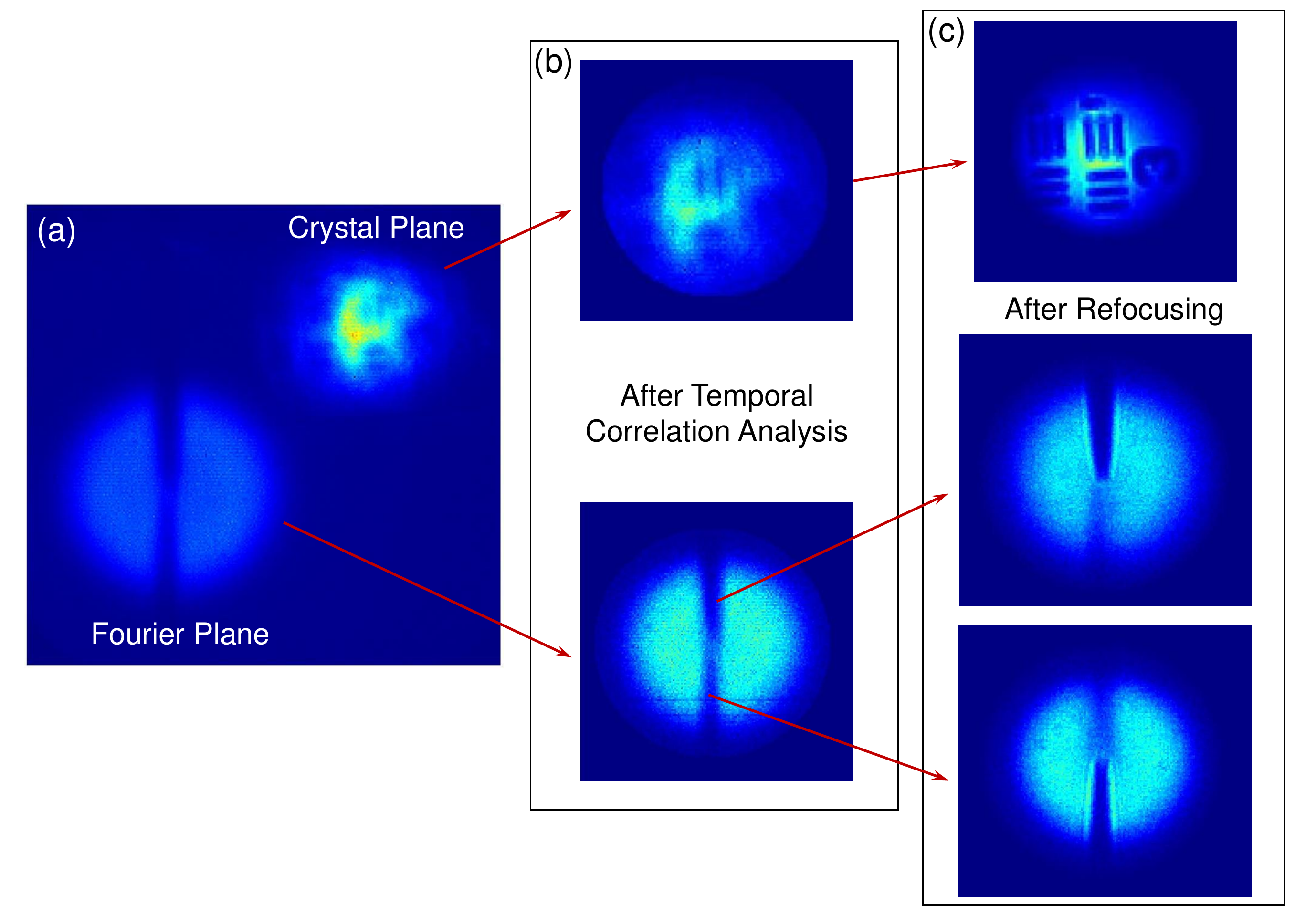}
    \caption{Demonstration of refocusing when sample is place in both the crystal plane and its Fourier plane. Here, a resolution target is placed out of focus in the crystal plane and two needles are placed at $\sim 10$\,cm on either side of the focus in the Fourier plane. (a) shows the raw image captured on camera. (b) shows the images after time correlation analysis. (c) shows the images after refocusing.}   
    \label{Refocus}
\end{figure}

\begin{figure}[h]
    \centering
    \includegraphics[width=1\linewidth]{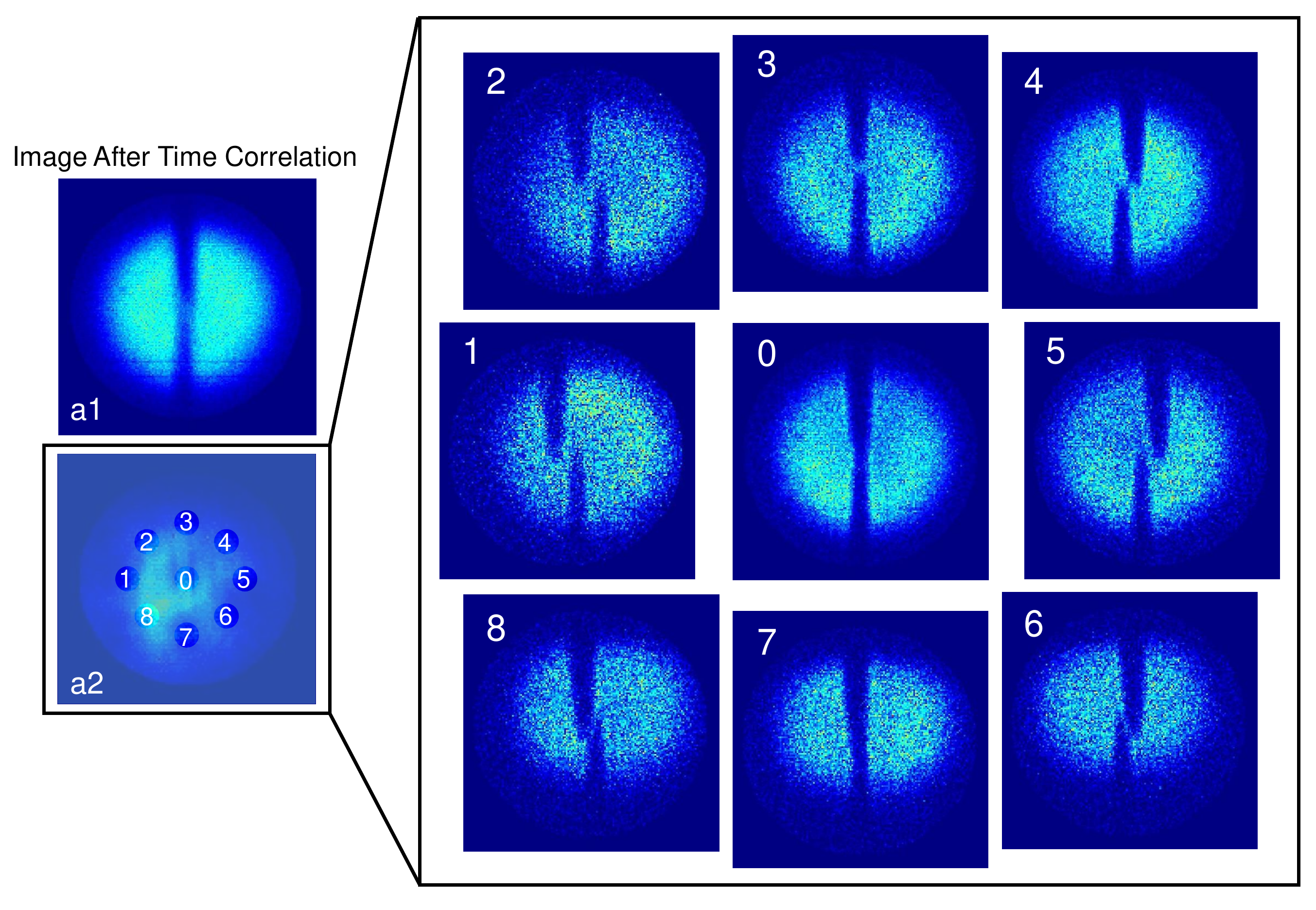}
    \caption{Images demonstrating parallax mapping and depth-of-field enhancement on 2 needles placed at different depth from the focus in the crystal Fourier plane. a1 and a2 shows the raw images taken in the Fourier plane and crystal plane respectively after time correlation analysis. To demonstrate parallax mapping, correlation events between the full Fourier plane and 9 sub-regions of the crystal plane (as indicated in a2) were selected out. Each sub-region of a2 corresponds to a different viewing angle of the needles, these are displayed as a set of 9 images, labeled 0-8, displayed on the right, each corresponding to the 9 labeled sub-regions (illumination/viewing angles) in a2 respectively.}
    \label{Parallax}
\end{figure}

In Figs.\,\ref{Refocus} and \ref{Parallax} we show the refocusing, parallax mapping and depth of field adjustment in QCRTI with the samples (2 needles) placed in the crystal's Fourier plane. As the SPDC photon beam is more collimated in this region, the needles can be placed much further away from the camera focus. The two needles used in Figs.\,\ref{Refocus} and \ref{Parallax} are placed at $\sim 10$\,cm on either side of the focus.

\bibliographystyle{unsrt}
\bibliography{QCRTIref}